\def\version{January 7, 2004}

\documentclass[reqno,11pt]{amsart}

\usepackage{epsf}

\def\be{\begin{equation}}
\def\ee{\end{equation}}
\def\ba{\begin{align}}
\def\bm{\begin{multline}}
\def\bfig{\begin{figure}[htb]}
\def\efig{\end{figure}}
\newcommand{\paper}[1]{{\it #1}, }
\newcommand{\journal}[4]{#1 #2, #3 (#4)}
\newcommand{\CMP}{Commun.\ Math.\ Phys.}
\newcommand{\HPA}{Helv.\ Phys.\ Acta}
\newcommand{\JSP}{J.\ Stat.\ Phys.}
\newcommand{\RMP}{Rev.\ Math.\ Phys.}
\numberwithin{equation}{section}
\newtheorem{theorem}{Theorem}%[section]
\newtheorem{proposition}[theorem]{Proposition}

\renewcommand{\thefootnote}{\arabic{footnote})}
\newcommand{\nn}{\nonumber}
\def\bbbone{{\mathchoice {\rm 1\mskip-4mu l} {\rm 1\mskip-4mu l} {\rm
1\mskip-4.5mu l} {\rm 1\mskip-5mu l}}}

\DeclareMathSymbol{\leqslant}{\mathalpha}{AMSa}{"36}
\DeclareMathSymbol{\geqslant}{\mathalpha}{AMSa}{"3E}
\DeclareMathSymbol{\doteqdot}{\mathalpha}{AMSa}{"2B}
\DeclareMathSymbol{\circlearrowright}{\mathalpha}{AMSa}{"08}
\DeclareMathSymbol{\subsetneq}{\mathalpha}{AMSb}{"28}
\DeclareMathSymbol{\supsetneq}{\mathalpha}{AMSb}{"29}
\renewcommand{\leq}{\;\leqslant\;}
\renewcommand{\geq}{\;\geqslant\;}

\newcommand{\dd}{{\rm d}}
\newcommand{\e}[1]{\,{\rm e}^{#1}\,}
\newcommand{\ii}{{\rm i}}
\newcommand{\sumtwo}[2]{\sum_{\substack{#1 \\ #2}}}

\def\dist{{\operatorname{dist\,}}}

\newcommand{\compl}{{\text{\rm c}}}

\newcommand{\const}{{\text{\rm const}}}
\newcommand{\upchi}{\raise 2pt \hbox{$\chi$}}
\def\writefig#1 #2 #3 {\rlap{\kern #1 truecm \raise #2 truecm
\hbox{#3}}}
\def\figtext#1{\smash{\hbox{#1}} \vspace{-5mm}}
\newcommand{\caF}{{\mathcal F}}

\newcommand{\bbQ}{{\mathbb Q}}

\newcommand{\bbZ}{{\mathbb Z}}

\newcommand{\bsrho}{{\boldsymbol \rho}}

\begin{document}

{\hfill\small\version}
\vspace{2mm}

\title{Segregation in the asymmetric Hubbard model}

\author{Daniel Ueltschi}

\address{Daniel Ueltschi \hfill\newline
Department of Mathematics \hfill\newline
University of Arizona \hfill\newline
Tucson, AZ 85721, USA \newline\indent
{\rm http://math.arizona.edu/$\sim$ueltschi}}
\email{ueltschi@math.arizona.edu}

\maketitle

\begin{quote}
{\small
{\bf Abstract.} We study the `asymmetric' Hubbard model, where hoppings of electrons
depend on their spin. For strong interactions and sufficiently asymmetric hoppings, it is
proved that the ground state displays phase separation away from half-filling. This
extends a recent result obtained with Freericks and Lieb for the Falicov-Kimball model. It
is based on estimates for the sum of lowest eigenvalues of the discrete Laplacian in
arbitrary domains.

\vspace{1mm}

}  % end \small

\vspace{1mm}
\noindent
{\footnotesize {\it Keywords:} Hubbard model; Falicov-Kimball model; phase separation.}

\vspace{1mm}
\noindent
{\footnotesize {\it 2000 Math.\ Subj.\ Class.:} 82B10, 82B20, 82B26}

\vspace{2mm}
\noindent
{\it Dedicated to Elliott Lieb on the occasion of his seventieth birthday.}
\end{quote}

\iffalse
\renewcommand{\thefootnote}{}
\footnote{\copyright{} 2003 by the author. This paper may be reproduced,
in its entirety, for non-commercial purposes.}
\setcounter{footnote}{0}
\renewcommand{\thefootnote}{\arabic{footnote}}
\fi

\section{Introduction}

Electronic properties of condensed matter are difficult to apprehend because of the
many-body interactions between quantum particles. It is necessary to consider simplified
models that capture the physics of various systems. Of great relevance is the Hubbard
model \cite{Hub} where spin-$\frac12$ electrons move on a lattice and interact via a local
Coulomb repulsion. Although a considerable simplification to the original problem, the
Hubbard model is still difficult to study and Hubbard himself considered an approximation
where particles of one spin are infinitely massive and behave classically.

The latter model was reinvented later by Falicov and Kimball in a different context,
namely in the study of the metal-semiconductor transition in rare-earth materials
\cite{FK}. Two species of electrons corresponding to different electronic bands are moving
on a lattice, and relevant interactions are between particles of different species.
Electrons carry spins but these turn out to be mathematically irrelevant and they can be
left aside. There exist many results for the Falicov-Kimball model. Let us mention proofs
of long-range order \cite{KL, BS, GJL, Ken, LM, MM, HK} and of phase separation
\cite{Ken2}; all these results are valid at half-filling, that is, the total density is
equal to 1. Interfaces were studied in \cite{DMN}. The ground state is segregated away
from half-filling \cite{FLU, FLU2}; see also \cite{Gol}. There exist reviews by Gruber and
Macris \cite{GM}, J\c edrzejewski and Lema\'nski \cite{JL}, and Freericks and Zlati\'c
\cite{FZ}. Less is known rigorously about the Hubbard model, see the survey by Lieb
\cite{Lieb2}.

We consider here a Hamiltonian that interpolates between Hubbard and Falicov-Kimball and
that describes two species of spinless fermions moving on $\bbZ^d$; particles have
different effective masses, and there is a local interaction involving particles of
different species. The Hamiltonian in second quantization is
\be
\label{defHam}
H_\Lambda = -\sumtwo{x,y \in \Lambda}{|x-y|=1} c_{x1}^\dagger c_{y1} - t \sumtwo{x,y \in
\Lambda}{|x-y|=1} c_{x2}^\dagger c_{y2} + U \sum_{x\in\Lambda} n_{x1} n_{x2}.
\ee
Here $\Lambda$ is a finite cube in $\bbZ^d$ and
$c_{xj}^\dagger$ and $c_{xj}$ are creation and annihilation operators of a fermion of
species $j$ at site $x$. The first two terms represent the kinetic energy of light and
heavy electrons respectively (we suppose that $0\leq t\leq1$). $n_{xj} = c_{xj}^\dagger
c_{xj}$ is a particle number operator. The positive parameter $U$  measures the strength
of the on-site repulsion between particles of species 1 and 2.

Setting $t=1$ yields the Hubbard model, and $t=0$ yields the Falicov-Kimball model. It is
interesting to note that the behavior of both models is similar when both particles have
density $\frac12$: for $d\geq2$, the ground state of the Hubbard model is a spin singlet
\cite{Lieb}, and the one of the Falicov-Kimball displays long-range order of the
chessboard type \cite{KL}. This holds for all strictly positive values of the coupling
constant $U$. It is natural to conjecture that long-range order occurs for all $t$.

Convergent perturbative expansions for large $U$ are a major source of results for the
Falicov-Kimball model, at least at half-filling. See \cite{DFFR} and \cite{KU} for general methods, and \cite{DFF} for a
discussion specifically to the Falicov-Kimball model. These methods are robust and extend to any
perturbation of the model. This holds in particular in the case of the asymmetric Hubbard model with
small $t$.

Our goal is to identify a phase with with segregation and to contrast it with chessboard
order and with high-temperature disorder. This suggests to look at the
following operator,
\be
\label{defcorr}
\sigma_\Lambda(x) = \frac1{|\Lambda|} \sum_{y\in\Lambda} [n_{y2} - n_{y+x,2}]^2.
\ee
The corresponding correlation function is given by the expectation of $\sigma_\Lambda(x)$
in the equilibrium state. We consider here the canonical ensemble where densities of light
and heavy particles are fixed to $\rho_1$ and $\rho_2$ respectively. High temperature
states are translation invariant and exponentially clustering, and the correlation function converges to
$2\rho_2 (1-\rho_2)$ as $x\to\infty$. Notice that $0 < 2\rho_2 (1-\rho_2) \leq \frac12$.
We identify here a domain of parameters where the expectation of $\sigma_\Lambda(x)$ is
zero in the ground state (segregation). At half-filling perturbation methods \cite{DFFR,KU,DFF} show
that it is close to 1 when $|x|$ is odd (chessboard order).

\begin{theorem} \hfill
\label{thmcorr}
Suppose that $\rho_1+\rho_2\neq1$. There exist $U_0<\infty$ and $t_0>0$ (that
depend on $\rho_1$ and $\rho_2$ only) such that for $U>U_0$ and $t<t_0$ we have
$$
\bigl( \Upsilon_\Lambda, \sigma_\Lambda(x) \Upsilon_\Lambda \bigr) =
O(|\Lambda|^{-\frac1d}).
$$
Here $\Upsilon_\Lambda$ is any ground state in the subspace where light and heavy
particles have densities $\rho_1$ and $\rho_2$, respectively.
\end{theorem}

This theorem extends the result of \cite{FLU,FLU2} for the Falicov-Kimball model.
Its proof proceeds by obtaining estimates for the ground state energy. The ground state is
a linear combination of states with a fixed configuration of heavy particles. The weight of
configurations with large `boundary' (pairs of nearest-neighbor sites where one is
occupied and one is empty) is small. Indeed, most of light particles are delocalized in the
remaining sites and their kinetic energy would
otherwise be great, as it is roughly proportional to the boundary (see \cite{FLU} and
Section \ref{secsum}). The pressure exerted by the light particles packs the heavy
particles together. The kinetic energy of heavy particles is therefore irrelevant, and
simple estimates suffice in bounding their contribution. These ideas are detailed in
Section \ref{secgsah}.

Section \ref{secsum} reviews the results for the sum of lowest eigenvalues of the discrete
Laplacian obtained in \cite{FLU}, with some improvements in the regime of low densities.
We discuss the segregated states of the asymmetric Hubbard model for all $0\leq t\leq1$ in Section \ref{secseg}.
For given densities of light and heavy particles, there is one free parameter to characterize
segregation: the proportion of volume occupied by each type of particles. The restricted phase diagram of
segregate states displays a transition between a phase where the local density of heavy particles
is maximum (that is, 1), and a phase where they have a local density that is strictly less than 1.
Section \ref{secgsah} is devoted to the proof of Theorem \ref{thmcorr}.
\\

{\it Acknowledgements:} I am grateful to both referees for useful comments, and to Lotfi
Hermi for drawing my attention to the paper \cite{Mel}. This paper is
dedicated to Elliott Lieb on the occasion of his seventieth birthday. With Tom Kennedy,
Elliott Lieb reinvented the Falicov-Kimball model and obtained the first rigorous
results \cite{KL}. I enormously benefitted from a collaboration with him
and Jim Freericks on segregation in this model; the present paper is directly
inspired by \cite{FLU}.

\section{Sum of lowest eigenvalues of the discrete Laplacian}
\label{secsum}

The sum of the $N$ lowest eigenvalues of the discrete Laplacian in a finite domain
$\Lambda \subset \bbZ^d$ gives the ground state energy of $N$ spinless, non-interacting
electrons hopping in $\Lambda$. This quantity is relevant to some problems of condensed
matter physics. Thermodynamics suggests that it is equal to a bulk term that is proportional to
the volume $|\Lambda|$ of the domain, plus a positive boundary correction that is
proportional to the boundary of $\Lambda$. Li and Yau proved in 1983 that the sum of
lowest eigenvalues of the continuum Laplacian is indeed bounded below by the bulk term
\cite{LY}. See \cite{LL}, Theorem 12.3, for a clear exposition. The proof readily adapts
to the case of the lattice.

The problem on the lattice turns out to be simpler and allows for bounds on the
boundary correction, for given `electronic density' $\rho = N/|\Lambda|$. Precisely, the
boundary correction can be bounded above and below by positive numbers times the `surface'
of the boundary. This was done in \cite{FLU}; this section contains some improvements in
the limit of low densities.

Corresponding statements in the continuum case have not been obtained yet. The best
statements seem to be the upper bound of Lieb and Loss, Theorem 12.11 in \cite{LL}, and
the lower bound of Melas \cite{Mel}, who obtained a positive correction of the order of
the size of the domain to the power $d-2$. However, these bounds are not proportional to
the boundary when the density is fixed.

For a finite domain $\Lambda \subset \bbZ^d$ the discrete Laplacian $h_\Lambda$ is defined
by
\be
\label{defham}
h_\Lambda \varphi(x) = -\sumtwo{y\in\Lambda}{|y-x|=1} \varphi(y) + 2d \varphi(x),
\ee
for all $x\in\Lambda$. Here $\varphi \in \ell_2(\Lambda)$ is a normalized, complex
function on $\Lambda$. If $\varphi$ is an eigenstate with eigenvalue $e$, so is
$(-1)^{|x|} \varphi$ with eigenvalue $4d-e$ (here $|x|$ denotes the $\ell_1$ norm of
$x\in\bbZ^d$). One also checks that $h_\Lambda\geq0$, and therefore its spectrum is
contained in $(0,4d)$ and is symmetric around $2d$. The bulk term involves the ground state
energy per site $e(\rho)$ of free fermions and it is given by
\be
e(\rho) = \frac1{(2\pi)^d} \int_{\varepsilon_k < \varepsilon_{\rm F}(\rho)} \varepsilon_k \, \dd k.
\ee
Here $k \in [-\pi,\pi]^d$ and $\varepsilon_k = 2d - 2\sum_{i=1}^d \cos k_i$. Notice that $|k|^2 -
\tfrac1{12} |k|^4 \leq \varepsilon_k \leq |k|^2$.
The `Fermi level' $\varepsilon_{\rm F}(\rho)$ is defined by the equation
\be
\rho = \frac1{(2\pi)^d} \int_{\varepsilon_k < \varepsilon_{\rm F}(\rho)} \dd k.
\ee

Let $B(\Lambda)$ denote the number of bonds connecting $\Lambda$ with its complement,
\be
B(\Lambda) = \bigl| \bigl\{ (x,y): x \in \Lambda, y \in \Lambda^\compl, \text{ and }
|x-y|=1 \bigr\} \bigr|.
\ee
If $S_{\Lambda,N}$ is the sum of the $N$ lowest eigenvalues of $h_\Lambda$, and $\rho =
N/|\Lambda|$ is the density, we are looking for bounds of the form
\be
\label{segrineq}
e(\rho) |\Lambda| + a(\rho) B(\Lambda) \leq S_{\Lambda,N} \leq e(\rho) |\Lambda| +
b(\rho) B(\Lambda)
\ee
with positive $a(\rho)$, $b(\rho)$, that are independent of the domain. It was proved
in \cite{FLU} that
\be
b(\rho) = \rho - \tfrac1{2d} e(\rho)
\ee
gives the optimal upper bound, that is saturated by domains consisting of isolated sites.
(The size of the boundary was defined differently in \cite{FLU} but minor changes in the proof
yield the upper bound stated here.)

We define $a(\rho)$ to be the minimal `surface energy' among all possible domains. Namely,
for $\rho \in [0,1] \cap \bbQ$,
\be
\label{defa}
a(\rho) = \inf_\Lambda \, \frac{S_{\Lambda,N} - e(\rho) |\Lambda|}{B(\Lambda)}.
\ee
The infimum is taken over all finite domains $\Lambda$ such that $\rho |\Lambda|
= N$ is an integer. The symmetry of the spectrum of $h_\Lambda$ around $2d$ implies that
$a(1-\rho) = a(\rho)$. We give below lower and upper bounds, stating in particular that
$a(\rho)>0$ for $0<\rho<1$. Many questions remain open, such as the existence of a
minimizer in \eqref{defa}; continuity of $a(\rho)$; monotonicity and convexity of
$a(\rho)$ for $0\leq\rho\leq\frac12$. It is even not clear whether the infimum
\eqref{defa} can be taken on connected sets. In order to state the bounds for $a(\rho)$,
let us introduce
\be
\label{defxi}
\xi(\rho) = \rho \, \varepsilon_{\rm F}(\rho) - e(\rho).
\ee

\begin{theorem}
\label{thmbornenergie}
For all $0<\rho\leq\frac12$, we have
$$
0 < a(\rho) \leq \tfrac1{2d} \xi(\rho).
$$
For small densities, we have
$$
a(\rho) >  \tfrac2{(3d)^3} \xi(\rho) \, \bigl( 1 - O(\rho^{2/d}) \bigr).
$$
\end{theorem}

We prove here that $a(\rho)$ is bounded
below by $\frac2{(3d)^3} \xi(\rho)$ at low densities and that it is smaller than
$\frac1{2d} \xi(\rho)$; notice that $\xi(\rho) \sim \rho^{1+\frac2d}$ as $\rho\to0$.
Efforts are made here to get the best possible factor. On the other
hand, pushing the range of densities instead, we could get a positive lower bound for
$0<\rho<(4\pi)^{-d/2} \Gamma(\frac d2+1)^{-1}$. Remaining densities are much more
difficult to treat and we refer to \cite{FLU} (and to \cite{Gol} for subsequent
improvements and simplifications).

\begin{proof}[Proof of the lower bound for $a(\rho)$ for low densities.]
We follow \cite{FLU}, with some improvements.
Let $\varphi_j$ be the eigenvector of $h_\Lambda$ corresponding to the $j$-th eigenvalue $e_j$, and
$\hat\varphi_j$ be its Fourier transform
\be
\hat\varphi_j(k) = \sum_{x\in\Lambda} \varphi_j(x) \e{\ii kx}, \quad k \in [-\pi,\pi]^d.
\ee
Then
\be
\label{expresS}
S_{\Lambda,N} = \frac1{(2\pi)^d} \int_{[-\pi,\pi]^d} \bsrho(k) \, \varepsilon_k \, \dd k,
\ee
where
\be
\label{bsrho}
\bsrho(k) = \sum_{j=1}^N |\hat\varphi_j(k)|^2 = |\Lambda| - \sum_{j=N+1}^{|\Lambda|}
|\hat\varphi_j(k)|^2.
\ee
We also observe that $\frac1{(2\pi)^d} \int \bsrho(k) \dd k = N$. One obtains a lower bound for
$S_{\Lambda,N}$ by taking the infimum of the right side of \eqref{expresS} over all positive functions $\bsrho$
smaller than $|\Lambda|$ and with the proper normalization. This gives the bulk term
\cite{LY,LL,FLU}. In order to extract the effect of the boundary, one strengthens the upper bound for
$\bsrho(k)$, aiming at $|\Lambda| - \const \cdot B(\Lambda)$. We start as in \cite{FLU} and write down a 
Schr\"odinger equation that is valid for all $x\in\bbZ^d$:
\be
-\sum_e \varphi_j(x+e) + 2d \varphi_j(x) + \upchi_{\Lambda^\compl}(x) \sum_{e: x+e \in
\Lambda} \varphi_j(x+e) = e_j \varphi_j(x).
\label{Schreq}
\end{equation}
It is understood that $\varphi_j(x)=0$ if $x\notin\Lambda$; the sums are over unit vectors $e$. The term
with the characteristic function $\upchi_{\Lambda^\compl}$ involves only sites that are close to the
boundary. The Fourier transform of this equation can be written as
\be
\label{Ftransf}
\varepsilon_k \hat\varphi_j(k) + (b_k,\varphi_j) = e_j \hat\varphi_j(k),
\ee
where $b_k$ is the following `boundary vector'
\be
\label{defb}
b_k(x) = \upchi_{\partial\Lambda}(x) \e{-\ii kx} \sum_{e:x+e\notin\Lambda} \e{-\ii ke}.
\ee
We introduced the set $\partial\Lambda$ of sites inside $\Lambda$ touching its complement
\be
\partial\Lambda = \{ x\in\Lambda: \dist(x,\Lambda^\compl)=1 \}.
\ee
We observe that $B(\Lambda) \leq \|b_k\|^2 \leq 2d B(\Lambda)$, the lower bound
holding at least when $|k|_\infty \leq \frac\pi4$. The last term of \eqref{bsrho} can then be
written using \eqref{Ftransf} as
\be
\label{labornesubtile}
\sum_{j=N+1}^{|\Lambda|} |\hat\varphi_j(k)|^2 = \sum_{j=N+1}^{|\Lambda|}
\frac{|(b_k,\varphi_j)|^2}{(e_j - \varepsilon_k)^2} \geq \frac{\bigl( \sum_{j=N+1}^{|\Lambda|} |e_j - \varepsilon_k| \, |(b_k,\varphi_j)|^2
\bigr)^4}{\bigl( \sum_{j=N+1}^{|\Lambda|} |e_j - \varepsilon_k|^2 \, |(b_k,\varphi_j)|^2 \bigr)^3}.
\ee
The lower bound follows from H\"older's inequality. One easily checks that
\be
\label{facile}
\begin{split}
&\sum_{j=N+1}^{|\Lambda|} |e_j - \varepsilon_k| \, |(b_k,\varphi_j)|^2 \geq (b_k, h_\Lambda b_k) -
(\varepsilon_k + e_N) \|b_k\|^2, \\
&\sum_{j=N+1}^{|\Lambda|} |e_j - \varepsilon_k|^2 \, |(b_k,\varphi_j)|^2 \leq (b_k, h_\Lambda^2 b_k) +
\varepsilon_k^2 \, \|b_k\|^2.
\end{split}
\ee
From now on we suppose $\varepsilon_k$ and $e_N$ to be small so that they add to less
than 1. 
Notice that $(b_k,h_\Lambda^2 b_k) = \| h_\Lambda b_k \|^2$. Because each site of
$\partial\Lambda$ has a neighbor outside $\Lambda$ and $b_k$ is zero there, we have
\be
\label{borneinf}
(b_k, h_\Lambda b_k) = \sum_{\{x,y\} : |x-y|=1} \bigl| b_k(x)-b_k(y) \bigr|^2
\geq \sum_{x\in\partial\Lambda} |b_k(x)|^2 = \| b_k \|^2.
\ee
Then \eqref{labornesubtile}, \eqref{facile}, and \eqref{borneinf}, imply that
\be
\label{uneborne}
\sum_{j=N+1}^{|\Lambda|} |\hat\varphi_j(k)|^2 \geq \frac{(b_k, h_\Lambda b_k)^4
(1-\varepsilon_k-e_N)^4}{(\|h_\Lambda b_k\|^2 + \varepsilon_k^2 \|b_k\|^2)^3}.
\ee
We estimate the denominator.
\be
\begin{split}
\|h_\Lambda b_k\|^2 &= \sum_{x\in\Lambda} \Bigl| \sumtwo{y\in\bbZ^d}{|y-x|=1} \bigl(
b_k(x)-b_k(y) \bigr) \Bigr|^2 \\
&\leq 2d \sum_{x\in\Lambda} \sumtwo{y\in\bbZ^d}{|y-x|=1} \Bigl| b_k(x)-b_k(y) \Bigr|^2\\
&= 2d \sumtwo{x,y \in \bbZ^d}{|x-y|=1} \Bigl| b_k(x)-b_k(y) \Bigr|^2 - 2d
\sumtwo{x\in\Lambda^\compl,y\in\Lambda}{|x-y|=1} |b_k(y)|^2 \\
&\leq 4d (b_k,h_\Lambda b_k) - 2d \|b_k\|^2.
\end{split}
\ee
Inserting this bound in \eqref{uneborne}, we obtain
\be
\sum_{j=N+1}^{|\Lambda|} |\hat\varphi_j(k)|^2 \geq (1-\varepsilon_k - e_N)^4 \|b_k\|^2
\frac{(b_k,h_\Lambda b_k) / \|b_k\|^2}{\bigl[ 4d - (2d-\varepsilon_k^2)
\frac{\|b_k\|^2}{(b_k,h_\Lambda b_k)} \bigr]^3}.
\label{unedeplus}
\ee
Simple analysis shows that the minimum of the fraction  under the condition $(b_k,h_\Lambda
b_k) \geq \|b_k\|^2$ is equal to $\frac2{(3d)^3} (1 - \frac{\varepsilon_k^2}{2d})$.
Furthermore, $b_k(x)$ is close to $b_0(x)$ for small $k$,
\be
\label{etunedeplus}
\begin{split}
|b_k(x)|^2 &= \Bigl| \sum_{e:x+e\notin\Lambda} (\cos ke - \ii \sin ke) \Bigr|^2 \geq \Bigl|
\sum_{e:x+e\notin\Lambda} \cos ke \Bigr|^2 \\
&\geq \Bigl| (1 - \tfrac12 \varepsilon_k)
\sum_{e:x+e\notin\Lambda} 1 \Bigr|^2 \geq (1 - \varepsilon_k) b_0^2(x).
\end{split}
\ee
Clearly, $(1-\varepsilon_k-e_N)^4 (1-\frac{\varepsilon_k^2}{2d}) (1-\varepsilon_k) \geq 1 - 6\varepsilon_k -
4e_N$. Using \eqref{unedeplus} and \eqref{etunedeplus} and since $\|b_0\|^2 \geq B(\Lambda)$, we get
\be
\sum_{j=N+1}^{|\Lambda|} |\hat\varphi_j(k)|^2 \geq \tfrac2{(3d)^3}
(1-6\varepsilon_k-4e_N) B(\Lambda).
\ee
We can insert this estimate into \eqref{bsrho} so as to get
\be
\label{enfin!}
\bsrho(k) \leq |\Lambda| - \tfrac2{(3d)^3} (1-6\varepsilon_k-4e_N) B(\Lambda).
\ee

Suppose we have a bound $\bsrho(k) \leq (1-\alpha) |\Lambda|$ for some $\alpha$ that is
independent of $k$. Lieb and Loss `bathtub principle' (Theorem 1.14 in \cite{LL}) yields
\be
\begin{split}
S_{\Lambda,N} &\geq (1-\alpha) |\Lambda| \frac1{(2\pi)^d} \int_{\varepsilon_k < \varepsilon_{\rm
F}(\frac\rho{1-\alpha})} \varepsilon_k \, \dd k \\
&= (1-\alpha) e(\tfrac\rho{1-\alpha}) |\Lambda|.
\end{split}
\ee
And because $(1-\alpha) e(\frac\rho{1-\alpha})$ is convex as a function of $\alpha$, and
that its derivative is equal to $\xi(\frac\rho{1-\alpha})$, we obtain
\be
S_{\Lambda,N} \geq e(\rho) |\Lambda| + \alpha \xi(\rho) |\Lambda|.
\ee

Let $0<\eta<1$. We define
\be
\alpha = \tfrac2{(3d)^3} (1-\eta) \tfrac{B(\Lambda)}{|\Lambda|}.
\ee
The condition \eqref{enfin!} implies that $\bsrho(k) \leq (1-\alpha)|\Lambda|$ for all $k$
such that $\varepsilon_k < \varepsilon_{\rm F}(\frac\rho{1-\alpha})$, provided the
following condition holds true,
\be
\label{unecondition}
\eta \geq 6\varepsilon_{\rm F}(\tfrac\rho{1-\alpha}) + 4e_N.
\ee
Given $\eta$, we restrict to densities small enough so that
\be
\label{autrecondition}
6\varepsilon_{\rm F} \bigl( \tfrac\rho{1-\frac4{27d^2}} \bigr) < \tfrac12 \eta.
\ee
(Notice that $\frac4{27d^2}$ is an upper bound for $\alpha$.) Then for all domains
$\Lambda$ and all numbers of electrons $N$ such that $4e_N < \frac12 \eta$,
the condition \eqref{unecondition} is satisfied and we obtain
\be
S_{\Lambda,N} \geq e(\rho) |\Lambda| + \tfrac2{(3d)^3} (1-\eta) \xi(\rho) B(\Lambda).
\ee

Consider now the case where \eqref{autrecondition} is fulfilled but $4e_N > \frac12 \eta$.
We define $N'$ such that $4e_{N'} \leq \frac12 \eta$ and $4e_{N'+1} > \frac12 \eta$. Then
\be
\begin{split}
S_{\Lambda,N} &= S_{\Lambda,N'} + \sum_{j=N'+1}^N e_j \\
&\geq e(\rho') |\Lambda| +
\tfrac2{(3d)^3} (1-\eta) \xi(\rho') B(\Lambda) + \tfrac18 \eta (\rho-\rho') |\Lambda|.
\end{split}
\ee
The right side is larger than $e(\rho) |\Lambda| + \frac2{(3d)^3} (1-\eta) \xi(\rho) B(\Lambda)$ provided that
\be
e(\rho') + \tfrac2{(3d)^3} (1-\eta) \xi(\rho') \tfrac{B(\Lambda)}{|\Lambda|} + \tfrac18
\eta (\rho-\rho') \geq e(\rho) + \tfrac2{(3d)^3} (1-\eta) \xi(\rho) \tfrac{B(\Lambda)}{|\Lambda|}.
\ee
It is enough to check that the function $\frac18 \eta \rho - e(\rho) - \frac4{27d^2}
\xi(\rho)$ is increasing. The derivative of $\xi(\rho)$ is equal to $\rho
\frac\dd{\dd\rho} \varepsilon_{\rm F}(\rho)$. It is possible to verify that
\be
\tfrac\dd{\dd\rho} \varepsilon_{\rm F}(\rho) < \tfrac{8\pi}d \Gamma(\tfrac d2+1)^{2/d}
\rho^{-1+\frac2d}
\ee
(the bound is optimal in the limit $\rho\to0$). The function above is therefore increasing
for small densities. The number $\eta$ can be chosen
arbitrarily small by taking the density small enough. Precisely, the condition is that
$\const \cdot \rho^{2/d} \leq \eta$. This means that given $\rho$, we can take $\eta =
O(\rho^{2/d})$.
\end{proof}

\begin{proof}[Proof of the upper bound for $a(\rho)$.]
Let $\Lambda$ be a (rather large) domain, and $\Lambda'$ be a set of isolated sites
outside of $\Lambda$. Let $r$ be such that $|\Lambda'| = r |\Lambda|$. The spectrum of
$h_{\Lambda\cup\Lambda'}$ is given by the union of the spectrum of $h_\Lambda$ and of
$\{2d\}$, the latter eigenvalue being at least $|\Lambda'|$ times degenerated.
We have $S_{\Lambda,N} \geq S_{\Lambda\cup\Lambda',N}$ (with equality if $\frac
N{|\Lambda|} \leq \frac12$)
and $B(\Lambda') = B(\Lambda) + 2dr |\Lambda|$. Using the upper bound for $S_{\Lambda,N}$
and the lower bound for $S_{\Lambda\cup\Lambda',N}$, we obtain (with $\rho = \frac
N{|\Lambda|(1+r)}$)
\be
e\bigl( (1+r) \rho \bigr) |\Lambda| + b\bigl( (1+r) \rho \bigr) B(\Lambda) \geq (1+r)
e(\rho) |\Lambda| + a(\rho) \bigl[ B(\Lambda) + 2dr |\Lambda| \bigr].
\ee
Reorganizing,
\be
a(\rho) \Bigl[ \frac{B(\Lambda)}{|\Lambda|} + 2dr \Bigr] \leq e\bigl( (1+r) \rho
\bigr) - (1+r) \, e(\rho) + b\bigl( (1+r) \rho \bigr) \frac{B(\Lambda)}{|\Lambda|}.
\ee
This inequality holds for any domain $\Lambda$ such that $(1+r) \rho |\Lambda|$ is an
integer. Ratios boundary/volume can be made
arbitrarily small and therefore the corresponding terms can be omitted. We obtain
\be
a(\rho) \leq \frac{e\bigl( (1+r) \rho \bigr) - (1+r) \, e(\rho)}{2dr}.
\ee
Taking the limit $r\to0$ yields the result.
\end{proof}

\section{A discussion of segregation}
\label{secseg}

Particles of different species segregate away from half-filling, at least for large $U$ and small $t$.
The domain splits into two subdomains, $\Lambda = \Lambda_1 \cup \Lambda_2$, with $\Lambda_1$
containing light particles only, and $\Lambda_2$ containing heavy particles only. This is true up to
boundary terms that do not contribute to the bulk energy. We neglect boundary terms in this section.

There is one free parameter that controls segregation, namely the ratio of the volumes occupied
by each phase. For $t=0$ the ground state is realized with $|\Lambda_2|=N_2$; it was argued in
\cite{FLU} that light particles exert a `pressure' that packs heavy particles together; this pressure
overcomes the tendency of heavy particles to delocalize so as to decrease their own kinetic energy. If
$t$ is large enough however, heavy particles will extend their domain. We study this mechanism in this
section, assuming that particles always segregate. From the point of view of rigorous results, we
obtain upper bounds for the ground state energy of the system.

\bfig
\epsfxsize=80mm
\centerline{\epsffile{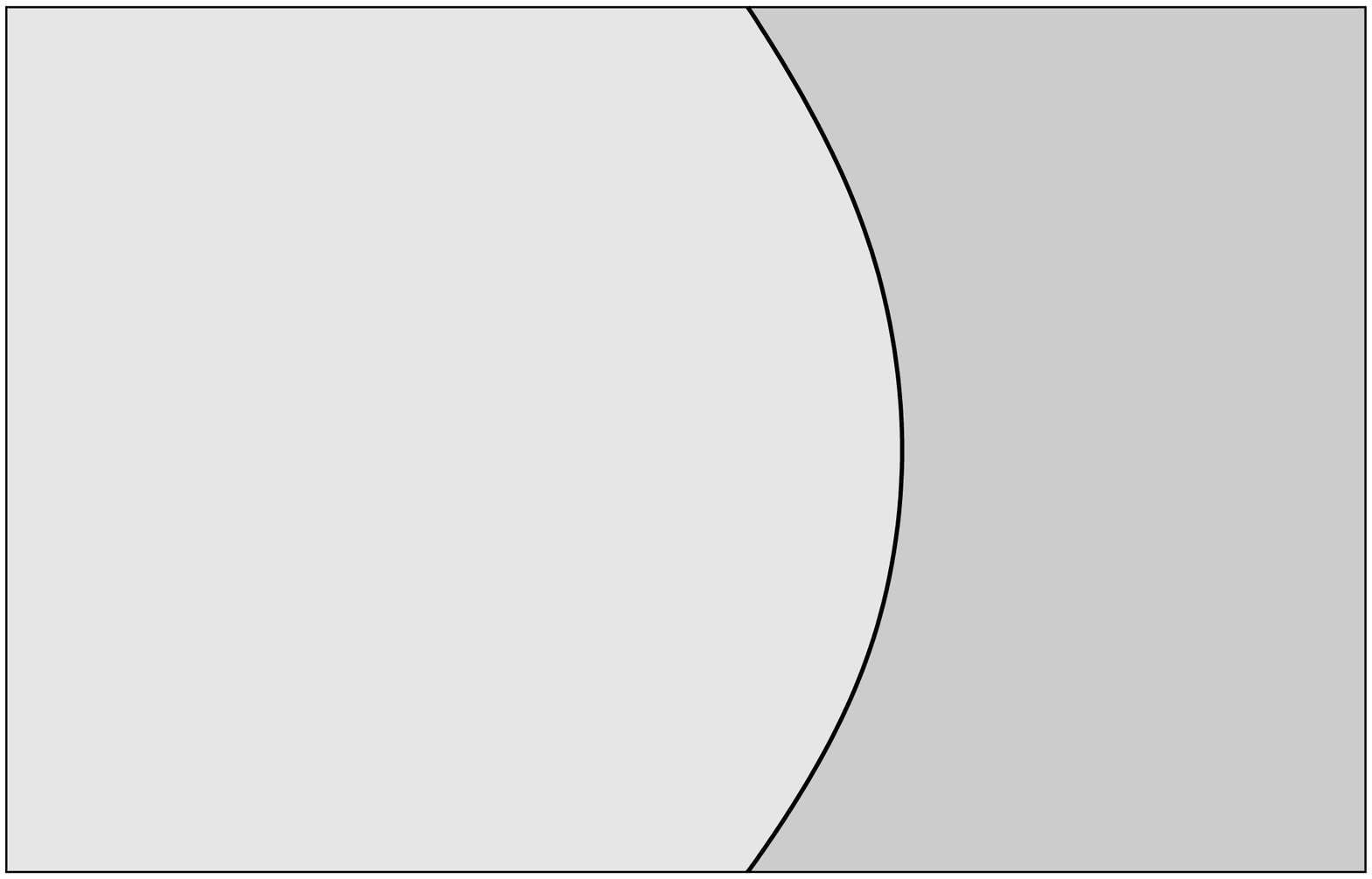}}
\figtext{
\writefig	6.15	3.8	{$\Lambda_1$}
\writefig	6.0	3.2	{$\frac{\rho_1}{1-\nu}$}
\writefig	9.9	2.6	{$\Lambda_2$}
\writefig	9.9	2.0	{$\frac{\rho_2}\nu$}
\writefig	11.4	0.8	{$\Lambda$}
}
\caption{A segregated state involves a partition of the domain into subdomains $\Lambda_1$ and
$\Lambda_2$ for light and heavy particles respectivey. The boundary between subdomains is supposed to
be small so that its contribution to the bulk energy is negligible. With $\nu$ such that $|\Lambda_2| =
\nu |\Lambda|$, densities inside each subdomain are $\frac{\rho_1}{1-\nu}$ and $\frac{\rho_2}\nu$.}
\label{figsegregation}
\end{figure}

We consider a finite domain $\Lambda \subset \bbZ^d$ partitioned in two subdomains $\Lambda_1$ and
$\Lambda_2$. We fix the number of particles $N_1$ and $N_2$ of light and heavy particles respectively,
and we denote the corresponding densities by $\rho_1 = N_1/|\Lambda|$ and $\rho_2 = N_2/|\Lambda|$. Let
$\nu = |\Lambda_2|/|\Lambda|$; we have $\rho_2 \leq \nu \leq 1-\rho_1$. Notice that the densities
inside each subdomain are $\frac{N_1}{|\Lambda_1|} = \frac{\rho_1}{1-\nu}$ and $\frac{N_2}{|\Lambda_2|} =
\frac{\rho_2}\nu$. Neglecting the contribution of boundaries, the energy per site of this
segregated state is
\be
e(\rho_1,\rho_2;\nu) = (1-\nu) \, e(\tfrac{\rho_1}{1-\nu}) + t \, \nu \, e(\tfrac{\rho_2}\nu).
\ee
For given densities $\rho_1$ and $\rho_2$ we are looking for the minimum of $e(\rho_1,\rho_2;\nu)$
with respect to $\nu$. One easily computes
\be
\frac\dd{\dd\nu} e(\rho_1,\rho_2;\nu) = \xi(\tfrac{\rho_1}{1-\nu}) - t \, \xi(\tfrac{\rho_2}\nu).
\ee
It is worth noticing that
$e(\rho_1,\rho_2;\nu)$ is convex in $\nu$, as its second derivative is positive ($\xi$ is
increasing). At $\nu=\rho_2$, we have
\be
\frac\dd{\dd\nu} e(\rho_1,\rho_2;\nu) \Big|_{\nu=\rho_2} = \xi(\tfrac{\rho_1}{1-\rho_2}) - 2dt.
\ee
This expression can be positive or negative, the critical parameter being $t_{\rm c} = \frac1{2d}
\xi(\frac{\rho_1}{1-\rho_2})$. On the other hand, the derivative of $e(\rho_1,\rho_2;\nu)$ at
$\nu=1-\rho_1$ is always positive (if $t\leq1$). Therefore the segregated state that has minimum
energy (among segregated states) is given as follows:
\begin{itemize}
\item If $t \leq t_{\rm c} = \frac1{2d} \xi(\frac{\rho_1}{1-\rho_2})$, the minimizer is $\nu =
\rho_2$, and the phase of heavy particles has density 1.
\item If $t_{\rm c} < t < 1$, the minimizer $\nu$ is between $\rho_2$ and
$\frac{\rho_2}{\rho_1+\rho_2}$, and the phase of heavy particles has a density strictly larger than
$\rho_1+\rho_2$ and strictly smaller than 1.
\item If $t=1$ the minimizer is $\nu = \frac{\rho_2}{\rho_1+\rho_2}$ and the phase of heavy particles
has density $\rho_1+\rho_2$.
\end{itemize}

This is illustrated in Fig.\ \ref{figphdseg}, that displays a restricted phase diagram where only
segregated states are considered.
This description is relevant only if a segregated state minimizes the energy. This is
proved in the case of small $t$.

A major open question in this model is {\it whether segregation really occurs for
$t>t_{\rm c}$}.

\bfig
\epsfxsize=90mm
\centerline{\epsffile{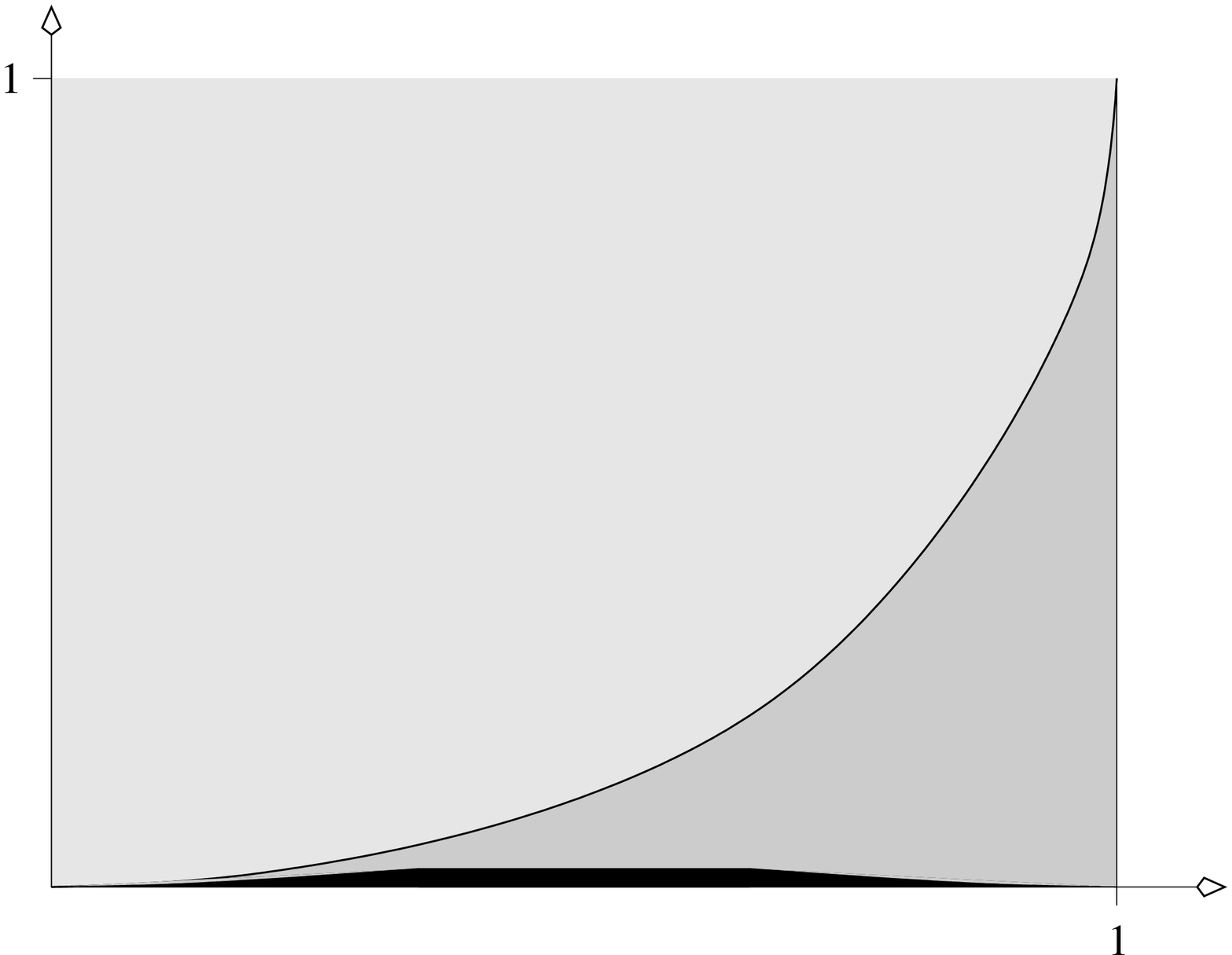}}
\figtext{
\writefig	2.9	7.4	{$t$}
\writefig	8.2	5.6	{$t_{\rm c} = \frac1{2d} \xi(\frac{\rho_1}{1-\rho_2})$}
\writefig	12.1	0.9	{$\frac{\rho_1}{1-\rho_2}$}
}
\caption{Restricted phase diagram for segregated states. The phase of heavy particles has density 1 in the
dark gray domain; its density is strictly less than 1 in the light gray domain. Segregation is proved
in the black domain when $U=\infty$ (and in a smaller domain when $U$ is large).}
\label{figphdseg}
\end{figure}

\section{The ground state of the asymmetric Hubbard model}
\label{secgsah}

Let $\caF(\Lambda)$ be the Fock space for spinless fermions in $\Lambda$. For $\Lambda' \subset \Lambda$, let
$\Phi_{\Lambda'} \in \caF(\Lambda)$ represents the state with $|\Lambda'|$ particles occupying all
sites of $\Lambda'$. $\{ \Phi_{\Lambda'} \}_{\Lambda' \subset \Lambda}$ is a basis for
$\caF(\Lambda)$. The state space for the asymmetric Hubbard model is $\caF(\Lambda) \otimes
\caF(\Lambda)$. Any function $\Upsilon_\Lambda \in \caF(\Lambda) \otimes \caF(\Lambda)$ can be written as
\be
\Upsilon_\Lambda = \sum_{\Lambda_1,\Lambda_2 \subset \Lambda} a_{\Lambda_1,\Lambda_2} \Phi_{\Lambda_1} \otimes
\Phi_{\Lambda_2},
\ee
with $\sum_{\Lambda_1,\Lambda_2} |a_{\Lambda_1,\Lambda_2}|^2 = 1$. Let
$a_{\Lambda_2} = (\sum_{\Lambda_1} |a_{\Lambda_1,\Lambda_2}|^2)^{1/2}$ and
$\Psi(\Lambda_2) \in \caF(\Lambda)$ be the normalized function such that
\be
a_{\Lambda_2} \Psi(\Lambda_2) = \sum_{\Lambda_1 \subset \Lambda} a_{\Lambda_1,\Lambda_2}
\Phi_{\Lambda_1}.
\ee
Then $\sum_{\Lambda_2} a_{\Lambda_2}^2 = 1$, and the function $\Upsilon_\Lambda$ can be written as
\be
\label{bonnerepresentation}
\Upsilon_\Lambda = \sum_{\Lambda_2 \subset \Lambda} a_{\Lambda_2} \Psi(\Lambda_2) \otimes
\Phi_{\Lambda_2}.
\ee

We derive in Proposition \ref{propcoeffgs} below an inequality for the coefficients
$a_{\Lambda_2}$ that will allow us to establish segregation in the ground state of the
strongly asymmetric Hubbard model.

Let $\caF(\Lambda;N)$ denote the Hilbert subspace of $\caF(\Lambda)$ corresponding to $N$ particles.
That is, it is spanned by $\{\Phi_{\Lambda'}\}$ with $|\Lambda'|=N$. All spaces $\caF(\Lambda;N_1)
\otimes \caF(\Lambda;N_2)$ are invariant under the action of $H_\Lambda$ since the latter conserves both
particle numbers. As before, we denote densities by $\rho_1 = \frac{N_1}{|\Lambda|}$ and $\rho_2 =
\frac{N_2}{|\Lambda|}$. The term $\gamma(U)$ that appears below was defined in \cite{FLU};
it behaves like $\frac{8d^2}U$ for large $U$.

\begin{proposition}
\label{propcoeffgs}
Let $\Upsilon_\Lambda$ be a ground state of $H_\Lambda$ in $\caF(\Lambda;N_1) \otimes
\caF(\Lambda;N_2)$. If $a(\frac{\rho_1}{1-\rho_2}) > \gamma(U) + t$, we have
$$
\sum_{\Lambda_2\subset\Lambda} a_{\Lambda_2}^2 B(\Lambda_2) \leq \frac{4d \, b(\frac{\rho_1}{1-\rho_2})
\rho_2^{1-\frac1d}}{a(\frac{\rho_1}{1-\rho_2}) - \gamma(U) - t} \;
|\Lambda|^{1-\frac1d}.
$$
\end{proposition}

\begin{proof}
We write the energy of a state using coefficients
$a_{\Lambda_2}$ defined above, and then use results obtained for the Falicov-Kimball model.
Let $T_\Lambda$ be the kinetic energy operator for particles in $\Lambda$; it acts on $\caF(\Lambda)$,
and can be written as
\be
T_\Lambda = -\sumtwo{x,y\in\Lambda}{|x-y|=1} c_x^\dagger c_y
\ee
where $c_x^\dagger$ and $c_x$ are creation and annihilation operators of a fermion at x. Notice that
the kinetic terms of \eqref{defHam} are given by $T_\Lambda \otimes \bbbone + t \bbbone \otimes
T_\Lambda$. Furthermore, for $\Lambda' \subset \Lambda$, let $V_{\Lambda,\Lambda'}$ be the operator
\be
V_{\Lambda,\Lambda'} = U \sum_{x\in\Lambda'} c_x^\dagger c_x.
\ee
It represents an external potential that is equal to $U$ on sites of
$\Lambda'$ and 0 otherwise. The energy of a state $\Upsilon_\Lambda$ given by
\eqref{bonnerepresentation} can be written as
\bm
\label{energie}
(\Upsilon_\Lambda, H_\Lambda \Upsilon_\Lambda) = t \sum_{\Lambda_2,\Lambda_2'} a_{\Lambda_2} a_{\Lambda_2'}
(\Psi(\Lambda_2),\Psi(\Lambda_2')) \, (\Phi_{\Lambda_2}, T_\Lambda \Phi_{\Lambda_2'}) \\
+ \sum_{\Lambda_2}
a_{\Lambda_2}^2 \bigl( \Psi(\Lambda_2), (T_\Lambda + V_{\Lambda,\Lambda_2})
\Psi(\Lambda_2) \bigr).
\end{multline}
(The sums are over sets satisfying $|\Lambda_2| = |\Lambda_2'| = N_2$.)
Notice that the first term of the right side is bounded by
\be
\label{bornecinetique}
t \sum_{\Lambda_2,\Lambda_2'}{}' a_{\Lambda_2} a_{\Lambda_2'} \leq t \Bigl(
\sum_{\Lambda_2,\Lambda_2'}{}'
a_{\Lambda_2}^2 \Bigr)^{1/2} \Bigl( \sum_{\Lambda_2,\Lambda_2'}{}' a_{\Lambda_2'}^2 \Bigr)^{1/2} = t
\sum_{\Lambda_2} a_{\Lambda_2}^2 B(\Lambda_2),
\ee
where the symbol $\sum{}'$ means a sum over pairs of sets $\Lambda_2, \Lambda_2' \subset \Lambda$ that
differ only by one site moved to a neigbor (that is, the symmetric difference $\Lambda_2 \triangle \Lambda_2'$ must be a pair
of nearest-neigbors).

The strategy of the proof is to consider the expression \eqref{energie} for the energy of the ground
state $\Upsilon_\Lambda$. We get an upper bound by using a trial function that is independent of the coefficients
$\{a_{\Lambda_2}\}$. We then estimate the second term of \eqref{energie} from below, using
the inequality \eqref{segrineq} for the segregation energy. The corresponding expression
involves the coefficients $\{a_{\Lambda_2}\}$ and we obtain the inequality stated in
Proposition \ref{propcoeffgs}.

Let $\Lambda_2\subset\Lambda$ be such that $|\Lambda_2|=N_2$, and let us consider
$\Psi(\Lambda_2)
\otimes \Phi_{\Lambda_2}$ where $\Psi(\Lambda_2)$ is a normalized function of $\caF(\Lambda,N_1)$
with support on $\Lambda_2^\compl$. We have
\ba
(\Psi(\Lambda_2) \otimes \Phi_{\Lambda_2}, H_\Lambda \, \Psi(\Lambda_2) \otimes \Phi_{\Lambda_2}) &=
(\Psi(\Lambda_2), T_{\Lambda_2^\compl} \Psi(\Lambda_2)) \nn\\
&\leq |\Lambda_2^\compl| e(\tfrac{\rho_1}{1-\rho_2}) + b(\tfrac{\rho_1}{1-\rho_2})
B(\Lambda_2).
\label{bornesup}
\end{align}
We used the upper bound in \eqref{segrineq}. We take for $\Lambda_2$ a square if possible,
or a domain with very close shape. Its boundary is less than $4d N_2^{1-\frac1d}$.
The boundary term in \eqref{bornesup} is then smaller than $4d
\, b(\tfrac{\rho_1}{1-\rho_2}) \rho_2^{1-\frac1d} |\Lambda|^{1-\frac1d}$. We use now the lower bound in \eqref{segrineq}. As stated in
this paper it holds only for $U=\infty$.
However, it was extended in \cite{FLU} to finite $U$; namely, it was shown there that
\be
\bigl( \Psi(\Lambda_2), (T_\Lambda + V_\Lambda) \Psi(\Lambda_2) \bigr) \geq
e(\tfrac{\rho_1}{1-\rho_2}) |\Lambda_2^\compl| + [a(\tfrac{\rho_1}{1-\rho_2}) - \gamma(U)]
B(\Lambda_2),
\ee
where $a(\cdot)$ is the minimal surface energy defined in \eqref{defa}.
Combining this with \eqref{energie} and \eqref{bornecinetique}, we
have for any ground state function $\Upsilon_\Lambda$,
\be
(\Upsilon_\Lambda, H_\Lambda \Upsilon_\Lambda) \geq -t \sum_{\Lambda_2} a_{\Lambda_2}^2 B(\Lambda_2) +
\sum_{\Lambda_2} a_{\Lambda_2}^2 \Bigl[ e(\tfrac{\rho_1}{1-\rho_2}) |\Lambda_2^\compl| +
\bigl( a(\tfrac{\rho_1}{1-\rho_2}) - \gamma(U) \bigr) B(\Lambda_2) \Bigr].
\ee
We now compare this expression with the upper bound \eqref{bornesup} and we obtain
\be
\sum_{\Lambda_2} a_{\Lambda_2}^2 B(\Lambda_2) \Bigl[ a(\tfrac{\rho_1}{1-\rho_2}) - \gamma(U)
- t \Bigr] \leq 4d \, b(\tfrac{\rho_1}{1-\rho_2})
\rho_2^{1-\frac1d} |\Lambda|^{1-\frac1d}.
\ee
\end{proof}

\begin{proof}[Proof of Theorem \ref{thmcorr}]
Using the decomposition \eqref{bonnerepresentation} for the ground state
$\Upsilon_\Lambda$, we have
\be
\label{correlations}
(\Upsilon_\Lambda, \sigma_\Lambda(x) \Upsilon_\Lambda) = \frac2{|\Lambda|} \sum_{y\in\Lambda} \sum_{\Lambda_2\subset\Lambda}
a_{\Lambda_2}^2 \upchi_{\Lambda_2}(y) \, \upchi_{\Lambda_2^\compl}(x+y).
\ee
It is clear that
\be
\label{bornecorr}
\sum_{y\in\Lambda} \upchi_{\Lambda_2}(y) \, \upchi_{\Lambda_2^\compl}(x+y) \leq B(\Lambda_2) \,
|x|_\infty,
\ee
and therefore
\be
(\Upsilon_\Lambda, \sigma_\Lambda(x) \Upsilon_\Lambda) \leq \frac{2|x|_\infty}{|\Lambda|} \sum_{\Lambda_2\subset\Lambda}
a_{\Lambda_2}^2 B(\Lambda_2).
\ee
Hole-particle symmetries in this model are similar to those in the Falicov-Kimball model, see \cite{KL},
and allow to restrict to the case $\rho_1+\rho_2<1$. We have $a(\frac{\rho_1}{1-\rho_2}) > \gamma(U) +
t$ if $U$ is large and $t$ is small, so that Proposition \ref{propcoeffgs} is valid.
We use it to control the sum above and we get Theorem \ref{thmcorr}.
\end{proof}

\end{document}